\documentclass[conference, 10pt]{IEEEtran}

\usepackage{amsmath,amssymb,epsfig,multicol}
\usepackage{epsf}

\begin{document}

\title{\huge{Multihop Diversity in Wideband OFDM Systems:}\\[0.5ex]
       \huge{The Impact of Spatial Reuse and Frequency Selectivity}} 

\author{\authorblockN{\"Ozg\"ur Oyman}
\authorblockA{
Intel Research\\
Santa Clara, CA 95054\\
Email: ozgur.oyman@intel.com}\and
\authorblockN{J. Nicholas Laneman}
\authorblockA{
University of Notre Dame\\
Notre Dame, IN 46556\\
Email: jnl@nd.edu}
}


%


\maketitle

\begin{abstract}
The goal of this paper is to establish which practical routing schemes for wireless networks are most suitable for wideband systems in the power-limited regime, which is, for example, a practically relevant mode of operation for the analysis of ultrawideband (UWB) mesh networks. For this purpose, we study the tradeoff between energy efficiency and spectral efficiency (known as the power-bandwidth tradeoff) in a wideband linear multihop network in which transmissions employ orthogonal frequency-division multiplexing (OFDM) modulation and are affected by quasi-static, frequency-selective fading. Considering open-loop (fixed-rate) and closed-loop (rate-adaptive) multihop relaying techniques, we characterize the impact of routing with spatial reuse on the statistical properties of the end-to-end conditional mutual information (conditioned on the specific values of the channel fading parameters and therefore treated as a random variable) and on the energy and spectral efficiency measures of the wideband regime. Our analysis particularly deals with the convergence of these end-to-end performance measures in the case of large number of hops, i.e., the phenomenon first observed in \cite{Oyman06b} and named as ``multihop diversity''. Our results demonstrate the realizability of the multihop diversity advantages in the case of routing with spatial reuse for wideband OFDM systems under wireless channel effects such as path-loss and quasi-static frequency-selective multipath fading.
\end{abstract}



%
\IEEEpeerreviewmaketitle

\section{Introduction}  

The design of large scale distributed networks (e.g., mesh and ad-hoc networks, relay networks) poses a set of new challenges to information theory, communication theory and network theory. Such networks are characterized by the large size of the network both in terms of the number of nodes (i.e., dense) and in terms of the geographical area the network covers. Each terminal can be severely constrained by its computational and transmission/receiving power. Moreover, delay and complexity constraints along with diversity-limited channel behavior may require transmissions under insufficient levels of coding protection causing link outages. These constraints require an understanding of the performance limits of such networks {\it jointly in terms of power and bandwidth efficiency and link reliability}, especially when designing key operational elements essential in these systems such as multihop routing algorithms, bandwidth allocation policies and relay deployment models. 

This paper applies tools from information theory and statistics to evaluate the end-to-end performance limits of various multihop routing algorithms in wireless networks focusing on the tradeoff between energy efficiency and spectral efficiency; which is also known as the {\it power-bandwidth tradeoff}. In particular, our main interest is in the power-limited {\it wideband} communication regime, in which transmitter power is much more costly than bandwidth. Since bandwidth is in abundance, communication in this regime is characterized by low signal-to-noise ratios (SNRs), very low signal power spectral densities, and negligible interference power. 

{\it Relation to Previous Work.} While the power-bandwidth tradeoff characterizations of various point-to-point and multi-user communication settings can be found in the literature, previous work addressing the fundamental limits over large adhoc wireless networks has generally focused either only on the energy efficiency performance \cite{Dana03} or only on the spectral efficiency performance \cite{Gupta00}\nocite{Gastpar02}-\cite{Boel2006}. The analytical tools to study the power-bandwidth tradeoff in the power-limited regime have been previously developed in the context of point-to-point
single-user communications \cite{Verdu02}, and were extended to multi-user (point-to-multipoint and multipoint-to-point) settings
\cite{Shamai01}-\nocite{Caire04}\nocite{Lapidoth03}\cite{Muharem03}, as well as to adhoc wireless networking examples of
single-relay channels \cite{Abbas03}-\cite{Yao03}, multihop networks under additive white Gaussian noise (AWGN) \cite{Laneman05} and dense multi-antenna relay networks \cite{oyman_pbt2005}. 

{\it Contributions.} This paper characterizes the power-bandwidth tradeoff in a wideband linear multihop network with quasi-static frequency-selective fading processes and orthogonal frequency-division multiplexing (OFDM) modulation over all links; with a key emphasis on the power-limited wideband regime. Our analysis considers open-loop (fixed-rate) and closed-loop (rate-adaptive) multihop relaying techniques and focuses on the impact of routing with spatial reuse on the statistical properties of the end-to-end conditional mutual information (conditioned on the specific values of the channel fading parameters and therefore treated as a random variable \cite{Ozarow94}) and on the energy and spectral efficiency measures of the wideband regime (computed from the conditional mutual information). Our results demonstrate the realizability of the multihop diversity advantages in the case of routing with spatial reuse for wideband OFDM systems under wireless channel effects such as path-loss and quasi-static frequency-selective multipath fading. The first author reported earlier analytical results for the case of multihop routing with no spatial reuse in \cite{Oyman06b}, which was the first work to observe the effect of multihop diversity for enhancing end-to-end link reliability in diversity-limited flat-fading wireless systems.

\section{Network Model and Definitions}

\subsection{General Assumptions} 

\begin{figure} [t]
\begin{center}
\includegraphics[width=3.4in, keepaspectratio]{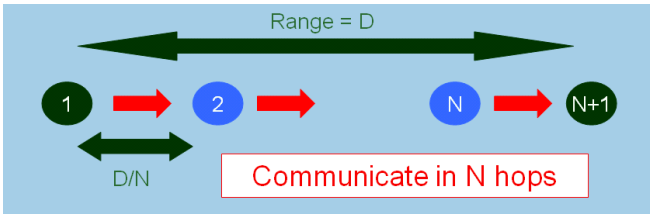}
\end{center}
\caption{Linear multihop network model.}
\label{linear_net}
\end{figure}

We model a linear multihop network as a network in which a pair of source and destination terminals communicate with each other by routing their data through multiple intermediate relay terminals, as depicted in Fig. \ref{linear_net}. If the linear multihop network consists of $N+1$ terminals; the source terminal is identified as ${\cal T}_1$, the destination terminal is identified as ${\cal T}_{N+1}$, and the intermediate relay terminals are identified as ${\cal T}_2$-${\cal T}_{N}$, where $N$ is the number of hops along the transmission path. Because terminals cannot transmit and receive at the same time in the same frequency band, we only focus on time-division based (half duplex) relaying, which orthogonalizes the use of the time and frequency resources
between the transmitter and receiver of a given radio. Moreover, we consider {\it full decoding} of the entire codeword at the intermediate relay terminals, which is also called {\it regeneration} or {\it decode-and-forward} in various contexts. In particular, for any given message to be conveyed from ${\cal T}_1$ to ${\cal T}_{N+1}$, we consider a simple $N$-hop decode-and-forward multihop routing protocol, in which, at hop $n$, relay terminal ${\cal T}_{n+1}, \, n=1,...,N-1$, attempts to fully decode the intended message based on its observation of the transmissions of terminal ${\cal T}_{n}$ and forwards its re-encoded version over hop $n+1$ to terminal ${\cal T}_{n+1}$. We consider multihop relaying protocols with no interference across different hops, as well as those with {\it spatial reuse}, for which we allow a certain number of terminals over the linear network to transmit simultaneously over the same time slot and frequency band. 

To facilitate parallel transmission of several packets through the linear multihop network, the available bandwidth is reused between transmitters, with a minimum separation of $K$ terminals between simultaneously transmitting terminals ($2 \leq K \leq N$); such that $N$ is divisible by $K$ and $M=N/K$ simultaneous transmissions are allowed at any time. Such spatial reuse schemes enable multiple nodes to transmit leading to more efficient use of
bandwidth, but introducing intra-route interference. For the case of no spatial reuse, we have $K=N$ and $M=1$. In decoding the message, terminals regard all interference signals not originating from the preceding node as noise, i.e., the receiver at terminal ${\cal T}_{n+1}$ treats all received signal components other than that from terminal ${\cal T}_n$ as noise.

\subsection{Channel and Signal Model} 

We consider the wideband channel over each hop to exhibit quasi-static frequency-selective fading with AWGN over the bandwidth of interest, and assume perfectly synchronized transmission/reception between the terminals. Using OFDM modulation turns the frequency-selective fading channel into a set of $W$ parallel frequency-flat fading channels rendering multi-channel equalization particularly simple since for each OFDM tone a narrowband receiver can be employed. We assume that the length of the cyclic prefix (CP) in the OFDM system is greater than the length of the discrete-time baseband channel impulse response. This assumption guarantees that the frequency-selective fading channel decouples into a set of parallel frequency-flat fading channels. Our channel model accommodates multihop routing protocols with spatial reuse as well as those without spatial reuse. At hop $n$ and tone $w$, the discrete-time memoryless complex baseband input-output channel relation is given by ($n=1,...,N$ and $w=1,...,W$)
$$
{y}_{n,w} =  \left( \frac{1}{d_{n}} \right)^{p/2} H_{n,w} \,s_{n,w} + \sum_{l \in {\cal L}_n} \left( \frac{1}{f_{n,l}} \right)^{p/2} G_{n,l,w} \,i_{n,l,w} + z_{n,w},
$$
where $y_{n,w} \in {\Bbb C}$ is the received signal at terminal ${\cal T}_{n+1}$,
$s_{n,w} \in {\Bbb C}$ is the temporally i.i.d. zero-mean
circularly symmetric complex Gaussian scalar transmit signal from ${\cal T}_n$
satisfying the average transmit power constraint ${\Bbb E}\left[|s_{n,w}|^2\right] = P_s$,
$i_{n,l,w} \in {\Bbb C}$ is the temporally i.i.d. zero-mean
circularly symmetric complex Gaussian scalar transmit signal from intra-route interference source $l$
satisfying the average transmit power constraint ${\Bbb E}\left[|i_{n,l,w}|^2\right] = P_i$,  
$z_{n,w} \in {\Bbb C}$ is the temporally white zero-mean circularly symmetric
complex Gaussian noise signal at ${\cal T}_{n+1}$, independent across $n$ and $w$ and independent from the input signals $\{s_{n,w}\}$ and $\{i_{n,l,w}\}$, with single-sided noise spectral density $N_0$, $d_{n}$ is the inter-terminal distance
between terminals ${\cal T}_n$ and ${\cal T}_{n+1}$, $f_{n,l}$ is the inter-terminal distance between interference source $l$ and terminal ${\cal T}_{n+1}$, set ${\cal L}_n$ contains the indices of the subset of terminals ${\cal T}_1$-${\cal T}_{N+1}$ over the linear multihop network contributing to the intra-route interference seen during the reception of terminal ${\cal T}_{n+1}$ and 
$p$ is the path loss exponent ($p \geq 2$). All of the discrete-time channels are assumed to be frequency-selective with $V$ delay taps indexed by $v=0,...,V-1$, under a certain power delay profile (PDP) such that their frequency responses sampled at tones $w=1,...,W$ are
$$
H_{n,w}=\sum_{v=0}^{V-1} h_{n,v} e^{-j 2 \pi v w / W},\,\,\,\,
G_{n,l,w}=\sum_{v=0}^{V-1} g_{n,l,v} e^{-j 2 \pi v w / W},
$$
for the signal and interference components, respectively, where $h_{n,v} \in {\Bbb C}$ and $g_{n,l,v} \in {\Bbb C}$ are random variables of arbitrary continuous distributions representing the signal and interference channel gains at receiving terminal ${\cal T}_{n+1}$, due to fading (including shadowing and microscopic fading effects) over the wireless links. We assume that the linear multihop network has a one-dimensional geometry such that the source terminal ${\cal T}_1$ and destination terminal ${\cal T}_{N+1}$ are separated by a distance $D$ and all intermediate terminals ${\cal T}_2$-${\cal T}_{N}$ (in that order) are equidistantly positioned on the line between ${\cal T}_1$ and ${\cal T}_{N+1}$, i.e., the inter-terminal distance $d_n$ is chosen as $d_n=D/N$. 

The channel fading statistics over the linear multihop network (modeled by random variables $\{h_{n,v}\}$ and $\{g_{n,l,v}\}$) are assumed to be based on i.i.d. realizations across different hops and taps (across $n$ and $v$). Furthermore, our channel model concentrates on the quasi-static regime, in which, once drawn, the channel variables $\{h_{n,v}\}$ and $\{g_{n,l,v}\}$ remain fixed for the entire duration of the respective hop transmissions, i.e., each codeword spans a single fading state, and that the channel coherence time is much larger than the coding block length, i.e., slow fading assumption. Although we assume that each receiving terminal ${\cal T}_{n+1}$ accurately estimates and tracks its channel and therefore possesses the perfect knowledge of the signal channel states $\{h_{n,v}\}_{v=0}^{V-1}$ and aggregate interference powers due to sources in ${\cal L}_n$, we consider two separate cases regarding the availability of channel state information (CSI) at the transmitters: 

(i) {\it Fixed-rate transmissions:} No terminal possesses transmit CSI which necessitates a fixed-rate transmission strategy for all terminals, where the rate is chosen to meet a certain level of reliability with a certain probability, 

(ii) {\it Rate-adaptive transmissions:} Each transmitting terminal ${\cal T}_{n},\,n=1,...,N$ possesses the knowledge of the channel states $\{h_{n,v}\}_{v=0}^{V-1}$ and aggregate interference powers due to sources in ${\cal L}_n$, and this allows for adaptively choosing the transmission rate over hop $n$ in a way that guarantees reliable communication provided that the coding blocklength is arbitrarily large.

It should be emphasized that we only assume the presence of local CSI at the terminals so that each terminal knows perfectly the receive (and possibly transmit) CSI regarding only its neighboring links, and our work does not assume the presence of global CSI at the terminals. In general, due to slow fading, each terminal in the linear multihop network may be able to obtain full channel state information (CSI) for its neighboring links through feedback mechanisms.

\subsection{Coding Framework}

To model block-coded communication over the linear multihop network, a $(\{M_n\}_{n=1}^N,\{Q_n\}_{n=1}^N,Q)$ multihop code ${\cal C}_Q$ is defined by a codebook of $\sum_{n=1}^N M_n$ codewords such that $M_n$ is the number of messages (i.e., number of codewords) for transmission over hop $n$, $Q_n$ is the coding blocklength over hop $n$, $R_n = (1/Q_n)\ln(M_n)$ is the rate of communication over hop $n$ (in nats per channel use), and $Q = \sum_{n=1}^N Q_n$ is the fixed total number of channel uses over the multihop link, representing a delay-constraint in the end-to-end sense, i.e., the $N$-hop routing protocol to convey each message from ${\cal T}_1$ to ${\cal T}_{N+1}$ takes place over the total duration of $\sum_{n=1}^N Q_n = Q$ symbol periods. Let ${\cal S}_{Q_n}$ be the set of all sequences of length $Q_n$ that can be transmitted on the channel over hop $n$ and ${\cal Y}_{Q_n}$ be the set of all sequences of length $Q_n$ that can be received. The codebook for multihop transmissions is determined by the encoding functions  $\phi_n,\,n=1,...,N$, that map each message $w_n\in {\cal W}_n = \{1,...,M_n\}$ over hop $n$ 
to a transmit codeword ${\bf s}_{n} \in {\Bbb C}^{W \times Q_n}$, where $s_{n,w} [q] \in {\cal S}_{1}$ is the transmitted
symbol over hop $n$ and tone $w$ during channel use $\sum_{m=1}^{n-1}Q_m+q,\,q=1,...,Q_n$. Each receiving terminal employs a decoding function $\psi_n \, , \,n=1,...,N$ to perform the mapping ${\Bbb C}^{W\times Q_n} \rightarrow \hat{w}_n \in {\cal W}_n$ based on its observed signal ${\bf y}_n \in {\Bbb C}^{W \times Q_n}$, where $y_{n,w}[q] \in {\cal Y}_{1}$ is the received symbol over hop $n$ and tone $w$ at time $\sum_{m=1}^{n-1}Q_m+q$. The codeword error probability for the $n$-th hop is given by $\epsilon_n = {\Bbb P}(\psi_n({\bf y}_n) \neq w_n)$. An $N$-tuple of multihop rates $(R_1,...,R_N)$ is
achievable if there exists a sequence of $(\{M_n\}_{n=1}^N, \{Q_n\}_{n=1}^N, Q)$
multihop codes $\{{\cal C}_Q:Q=1,2,...\}$ with $Q=\sum_{n=1}^N Q_n$, $Q_n >0, \forall n$, and vanishing $\epsilon_n,\, \forall n$.

\subsection{Power-Bandwidth Tradeoff Measures} 

We assume that the linear multihop network is supplied with finite total average transmit power $P$ (in Watts (W)) over unconstrained bandwidth $B$ (in Hertz (Hz)). The available transmit power is shared equally among $M=N/K$ simultaneous transmissions and $W$ OFDM tones of equal bandwidth $B/W$, leading to $P_s=P/(M\,W)$ and $P_i=P/(M\,W)$. If the transmitted codewords over the linear multihop network are chosen to achieve the desired end-to-end data rate per unit bandwidth (target spectral efficiency) $R$, reliable communication requires that $R \leq {\cal I}\left({E_b}/{N_0}\right)$ as $Q_n \rightarrow \infty,\,\forall n$, where ${\cal I}$ denotes the conditional mutual information (in nats/second/Hertz (nats/s/Hz)) which is a random variable under quasi-static fading, and $E_b/N_0$ is the energy per information bit normalized by the background noise spectral level, expressed as $E_b/N_0 = \mathsf{SNR}/I(\mathsf{SNR})$ for $\mathsf{SNR} = P/(N_0B)$ and $I$ denoting the conditional mutual information as a function of $\mathsf{SNR}$ \footnote{The use of $I$ and ${\cal I}$ avoids assigning the same symbol to conditional mutual information functions of $\mathsf{SNR}$ and ${E_b}/{N_0}$.}. 

There exists a tradeoff between the efficiency measures ${E_b}/{N_0}$ and ${\cal I}$ (known as the power-bandwidth tradeoff) in achieving a given target data rate. When ${\cal I} \ll 1$, the system operates in the {\it power-limited wideband regime}; i.e., the bandwidth is large and the main concern is the limitation on power. Particular emphasis throughout our analysis is placed on this wideband regime, i.e., regions of low ${E_b}/{N_0}$.

Defining $({E_b}/{N_0})_{\mathrm{min}}$ as the minimum system-wide
${E_b}/{N_0}$ required to convey any positive rate reliably, we have 
$({E_b}/{N_0})_{\mathrm{min}} = \min_{\vspace*{2mm}\mathsf{SNR}} \, \mathsf{SNR}/I(\mathsf{SNR})$.
In most scenarios, ${E_b}/{N_0}$ is minimized in the wideband regime when $\mathsf{SNR}$ is low and $I$ is near zero.
We consider the first-order behavior of $\cal{I}$ as a function of
${E_b}/{N_0}$ when ${\cal I} \rightarrow 0$ by analyzing the affine function (in decibels)
\footnote{ $\,\,u(x)=o(v(x)), x \rightarrow L$ stands for $\lim_{x
\rightarrow L}\frac{u(x)}{v(x)}=0$.} \footnote{ $\, \, \, =^{\!\!\!\!\!\mbox{\tiny
a.s.}}\,\,\,$ denotes statistical equality with probability $1$.} 
$$
10\log_{10}\frac{E_b}{N_0} \left( {\cal I} \right) \, \, \, =^{\!\!\!\!\!\mbox{\tiny
a.s.}}\,\,\, 10\log_{10}\frac{E_b}{N_0}_{\mathrm{min}} +
\frac{\cal I}{S_0}10\log_{10}2 + o({\cal I}),
$$
where $S_0$ denotes the ``wideband'' slope of mutual
information in b/s/Hz/(3 dB) at the point $({E_b}/{N_0})_{\mathrm{min}}$, 
$$
S_0 \, \, \, =^{\!\!\!\!\!\mbox{\tiny
a.s}}\,\,\, \lim_{\frac{E_b}{N_0} \downarrow \frac{E_b}{N_0}_{\mathrm{min}}}
\frac{{\cal I}(\frac{E_b}{N_0})}{10\log_{10}\frac{E_b}{N_0}-10\log_{10}
\frac{E_b}{N_0}_{\mathrm{min}}}10\log_{10}2 .
$$
It can be shown that \cite{Verdu02} 
\begin{equation}
\frac{E_b}{N_0}_{\mathrm{min}} \, \, \, =^{\!\!\!\!\!\mbox{\tiny
a.s}}\,\,\, \lim_{\mathsf{SNR} \rightarrow 0} \,\frac{\ln{2}}{\dot{I}(\mathsf{SNR})},
\label{min_ener}
\end{equation}
and
\begin{equation}
S_0 \, \, \, =^{\!\!\!\!\!\mbox{\tiny
a.s}}\,\,\, \lim_{\mathsf{SNR} \rightarrow 0} \frac{2{\left[ \dot{I}(\mathsf{SNR}) \right]}^2}{-\ddot{I}(\mathsf{SNR})},
\label{wideband_slope}
\end{equation}
where $\dot{I}$ and $\ddot{I}$ denote the first and second order derivatives of
$I(\mathsf{SNR})$ (evaluated in nats/s/Hz) with respect to $\mathsf{SNR}$.

\section{Power-Bandwidth Tradeoff in Wideband Linear Multihop Networks}

\begin{figure} [t]
\begin{center}
\includegraphics[width=3.4in, keepaspectratio]{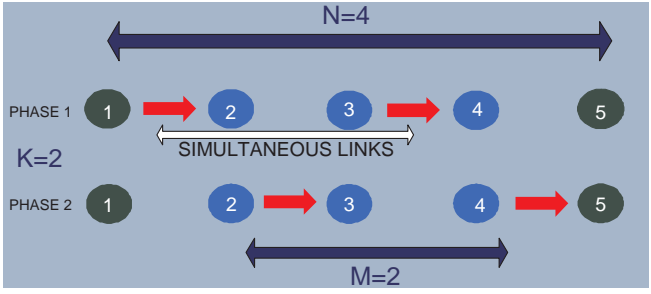}
\end{center}
\caption{Linear multihop network model with spatial reuse and time-sharing ($N=4, K=2, M=2$).}
\label{linear_net_reuse}
\end{figure}

We begin this section by characterizing the end-to-end mutual information over the linear multihop network considering the use of point-to-point capacity achieving codes over each hop. For the mutual information analysis, we do not impose any delay constraints on the multihop system and allow each coded transmission to have an arbitrarily large blocklength (i.e. assume large $\{Q_n\}$), although we will be concerned with the relative sizes of blocklengths over multiple hops. It is assumed that the nodes share a band of radio frequencies allowing for a signaling rate of $B$ complex-valued symbols per second. For any given spatial reuse separation $K$, the time-division based multihop routing protocol is specified by the time-sharing constants $\{\lambda_k\}_{k=1}^K,\,\sum_{k=1}^K \lambda_k=1$, where $\lambda_{k} \in [0,1]$ is defined as the fractional time during which reuse phase $k$ is active ($k=1,...,K$), with simultaneous transmission and reception over the corresponding $M=N/K$ hops. An example time-division based multihop routing protocol with spatial reuse and time-sharing is depicted in Fig. \ref{linear_net_reuse} for $N=4,K=2,M=2$. 

For any given reuse phase $k$, the set of hops performing simultaneous transmissions are indexed by $m=1,...,M$. If the transmitted codewords over reuse phase $k$ are chosen based on the data rate per unit bandwidth (spectral efficiency) $\tilde{R}_k$, reliable communication requires that the condition $\tilde{R}_k \leq \min_m I_{k,m}(\mathsf{SNR})$ is met for all $k$, where $I_{k,m}$ denotes the mutual information over link $m$ during reuse phase $k$; such that hop index $n=(m-1)K+k$. The end-to-end conditional (instantaneous) mutual information $I$ of the linear multihop network
can be expressed in the form \cite{Oyman06b, Laneman05}
\begin{equation}
I(\mathsf{SNR}) = \max_{\sum_{k=1}^K \lambda_k = 1} \min_k \left\{ \lambda_k 
\min_{m} I_{k,m}(\mathsf{SNR}) \right\},
\label{cap_minimax}
\end{equation}
where $I_{k,m}(\mathsf{SNR})$ is the conditional mutual information given (in nats/s/Hz) by \cite{Boel_cap99}
\begin{equation}
I_{k,m}(\mathsf{SNR}) = \frac{1}{W} \sum_{w=1}^W \ln\left(1 + \mathsf{SINR}_{(m-1)K+k,w}(\mathsf{SNR})\right),
\label{cap_fsh}
\end{equation}
as a function of the received signal-to-interference-and-noise ratio (SINR), which is given at terminal ${\cal T}_{n+1}$ and tone $w$ by
$$
\mathsf{SINR}_{n,w}(\mathsf{SNR}) =  \frac{N^{p-1} \,K \, |H_{n,w}|^2\,\mathsf{SNR}}{D^{p}} \left(1+\zeta_{n,w}(\mathsf{SNR})\right)^{-1},
$$
where $\zeta_{n,w}(\mathsf{SNR})$ is the aggregate intra-route interference power scaled down by noise power that satisfies $\lim_{\mathsf{SNR} \rightarrow 0} \zeta_n(\mathsf{SNR})=0$.

\subsection{Fixed-Rate Multihop Relaying}

A suboptimal strategy that yields a lower bound to the conditional mutual information in (\ref{cap_minimax}) is equal time-sharing ($\lambda_k = 1/K$) and fixed-rate (open-loop) transmission over all hops, i.e. the rate over hop $n$ equals $R_n=R,\,\forall n$ for some fixed value of $R$. This strategy is applicable in the absence of rate adaptation mechanisms if CSI is not available at the transmitters. In this setting, the end-to-end conditional mutual information can be expressed as
\begin{eqnarray}
I(\mathsf{SNR}) & = & \frac{1}{K\,W} \, \min_{k,m} \, \sum_{w=1}^W \ln \left(1 + \mathsf{SINR}_{(m-1)K+k,w}(\mathsf{SNR}) \right) \nonumber \\
&=& \frac{1}{K\,W} \, \min_n \, \sum_{w=1}^W \ln \left(1 + \mathsf{SINR}_{n,w}(\mathsf{SNR}) \right).
\label{cap_eq_ts}
\end{eqnarray}

{\bf \noindent Theorem 1:} {\it In the wideband regime, for time-division based linear
multihop networks employing the fixed-rate decode-and-forward relaying protocol (equal time-sharing),
the power-bandwidth tradeoff can be characterized as a function of
the channel fading parameters through the following relationships:}
$$
\frac{E_b}{N_0}_{\mathrm{min}} \, \, =^{\!\!\!\!\!\mbox{\tiny
a.s.}}\,\,\, \frac{\ln{2}}{\min_n\,(1/W) \sum_{w=1}^W |H_{n,w}|^2} \left(\frac{D^p}{N^{p-1} \,K}\right),
$${\it and}
$$
S_0 \, \, \, =^{\!\!\!\!\!\mbox{\tiny
a.s.}}\,\,\, \frac{2}{K}.
$$
{\it In the limit of large $N$, $(E_b/N_0)_{\mathrm{min}}$ converges in distribution as follows:}\footnote{ $\, \, \, \longrightarrow^{\!\!\!\!\!\!\!\mbox{\tiny
d}}\,\,\,$ denotes convergence in distribution.}
$$
\frac{E_b}{N_0}_{\mathrm{min}} \, \longrightarrow^{\!\!\!\!\!\!\!\mbox{\tiny
d}}\,\,\,\,\,\,\,\, \frac{\ln{2}}{a_N \, \Theta + b_N} \left(\frac{D^p}{N^{p-1} \,K}\right),
$$
{\it where $a_N>0$, $b_N$ are sequences of constants and $\Theta$ follows one of the three families of extreme-value distributions $\mu$: i) Type I, $\mu(x) = 1 - \exp \left( -\exp(x) \right),\,-\infty<x<\infty,$ ii) Type II,   
$\mu(x) = 1 - \exp \left( -(-x)^{-\gamma} \right),\,\gamma >0$ if $x<0$ and
$\mu(x) = 1$ otherwise, iii) Type III, $\mu(x) = 1 - \exp \left( -x^{\gamma} \right),\,\gamma >0$ if $x \geq 0$ and
$\mu(x) = 0$ otherwise.}

\vspace*{2mm}

{\noindent \bf Proof.} We begin by applying (\ref{min_ener})-(\ref{wideband_slope}) to (\ref{cap_eq_ts}), which yields the non-asymptotic results of the theorem. Denoting $\beta_N = \min_{n=1,...,N} \,(1/W) \sum_{w=1}^W |H_{n,w}|^2$, if there exist sequences of constants $a_N > 0$, $b_N$, and some nondegenerate distribution function $\mu$ such that $(\beta_N - b_N)/a_N$ converges in distribution to $\mu$ as $N \rightarrow \infty$, i.e.,
$$
{\Bbb P}\left(\frac{\beta_N - b_N}{a_N} \leq x \right) \longrightarrow \mu(x)\,\,\,\,\,\,\mathrm{as} \,\, N \rightarrow \infty,
$$
then $\mu$ belongs to one of the three families of extreme-value distributions above \cite{Leadbetter83}. The exact asymptotic limiting distribution is determined by the distribution of $(1/W) \sum_{w=1}^W |H_{n,w}|^2$, and to which one of the three domains of attraction it belongs. Consequently, we have $\beta_N \, \longrightarrow^{\!\!\!\!\!\!\!\mbox{\tiny d}}\,\,\,\, a_N \,\, \Theta + b_N$, which completes the proof of the theorem. \hfill \fbox

\vspace*{2mm}

In the presence of non-ergodic, or even ergodic but slow fading channel variations, one approach toward the
information-theoretic characterization of the end-to-end performance under fixed-rate transmissions (in the absence of transmit CSI at all terminals) involves the consideration of
{\it outage probability} \cite{Ozarow94}. We define the end-to-end outage in a linear multihop network as the event that
the conditional mutual information based on the instantaneous channel fading parameters $\{h_{n,v}\}$ and $\{g_{n,l,v}\}$ cannot support the considered data rate. 
Expressed mathematically, the end-to-end outage probability is given in terms of end-to-end conditional mutual information $I(\mathsf{SNR})$ as
$P_{\mathrm{out}} = {\Bbb P} \left( I(\mathsf{SNR}) < R \right)$,
where $R$ is the desired end-to-end data rate per unit bandwidth (spectral efficiency). Following the results of Theorem 1, a similar outage characterization is applicable to the power-bandwidth tradeoff in the wideband regime; in particular we can write $(E_b/N_0)_{\mathrm{min}}$ as 
$$
\frac{E_b}{N_0}_{\mathrm{min,out}} = \frac{\ln{2}}{a_N \,\mu^{-1}(P_{\mathrm{out}})  + b_N} \left(\frac{D^p}{N^{p-1} \,K}\right).
$$

\subsection{Rate-Adaptive Multihop Relaying}

The conditional mutual information in (\ref{cap_minimax}) is achievable by the linear
multihop network under optimal time-sharing and rate adaptation to instantaneous fading variations. Because the transmission rate of each codeword over each hop is chosen so that reliable decoding is always possible (the rate is changed on a codeword by codeword basis to adapt to the instantaneous rate which depends on the channel fading conditions), the system is never in outage under this closed-loop strategy (assuming infinite block-lengths). Although outage may be irrelevant on a per-hop basis (full reliability given infinite block-lengths), investigating the statistical properties of end-to-end mutual information over the linear multihop network still yields beneficial insights in applications sensitive to certain QoS constraints (e.g. throughput, reliability, delay or energy constraints). Applying Lemma 1 in \cite{Oyman06b}, the end-to-end conditional mutual information under the rate-adaptive multihop relaying strategy becomes
\begin{equation}
I(\mathsf{SNR}) = \left(\sum_{k=1}^K \frac{1}{\min_{m} I_{k,m}(\mathsf{SNR})}\right)^{-1},
\label{cap_optimal_ts}
\end{equation}
where $I_{k,m}(\mathsf{SNR})$ was given earlier in (\ref{cap_fsh}).

\vspace*{2mm}

{\bf \noindent Theorem 2:} {\it In the wideband regime, for time-division based linear
multihop networks employing the rate-adaptive decode-and-forward relaying protocol (optimal time-sharing),
the power-bandwidth tradeoff can be characterized as a function of
the channel fading parameters through the following relationships:}
$$\frac{E_b}{N_0}_{\mathrm{min}} \, \, =^{\!\!\!\!\!\mbox{\tiny
a.s}}\,\,\, \left(\frac{D^p}{N^{p-1} \,K}\right)
\sum_{k=1}^K \frac{\ln{2}}{\min_m (1/W) \sum_{w=1}^W |H_{(m-1)K+k,w}|^2}, 
$$
{\it and}
$$
S_0 \, \, \, =^{\!\!\!\!\!\mbox{\tiny
a.s.}}\,\,\, \frac{2}{K}.
$$
{\it In the limit of large $N$ and for fixed $M$, $(E_b/N_0)_{\mathrm{min}}$ converges almost surely (with probability $1$) to the deterministic quantity} \footnote{ $\, \, \, \longrightarrow^{\!\!\!\!\!\!\!\!\mbox{\tiny
a.s.}}\,\,\,$ denotes convergence with probability $1$.}
$$
\frac{E_b}{N_0}_{\mathrm{min}} \, \, \longrightarrow^{\!\!\!\!\!\!\!\!\mbox{\tiny
a.s.}}\,\,\, \ln{2}\left(\frac{D^{p}}{N^{p-1}}\right) \chi +o\left(\frac{1}{N^{p-1}}\right).
$$
{\it where the constant $\chi$ is given by}
$$
\chi = {\Bbb
E}\left[ \frac{1}{\min_{m=1,...,M}(1/W)\sum_{w=1}^W|H_{m,w}|^2} \right].
$$
\vspace*{.1mm}

\subsection{Remarks on Theorems 1 and 2} 
 
Theorems 1 and 2 suggest that the channel dependence of the power-bandwidth tradeoff is reflected by the randomness of $(E_b/N_0)_{\mathrm{min}}$ for both fixed-rate and rate-adaptive multihop relaying schemes in the presence of spatial reuse and frequency selectivity. We observe under rate-adaptive relaying in the wideband regime that, as the number of hops tends to infinity, $(E_b/N_0)_{\mathrm{min}}$ converges almost surely to a \emph{deterministic} quantity independent of the fading channel realizations. Similarly, for fixed-rate relaying, we observe a weaker convergence (in distribution) for $(E_b/N_0)_{\mathrm{min}}$ in the case of asymptotically large number of hops. This averaging effect achieved by fixed-rate and rate-adaptive relaying schemes can be interpreted as {\it multihop diversity}, a phenomenon first observed in \cite{Oyman06b} for routing with no spatial reuse in frequency-flat fading channels; and now shown to be also realizable with spatial reuse and frequency selectivity. Although fixed-rate relaying for asymptotically large $N$ improves the outage performance, this framework does not yield the fast averaging effect that leads to the strong convergence of $(E_b/N_0)_{\mathrm{min}}$, that is observed under rate-adaptive relaying. However, the variability of $(E_b/N_0)_{\mathrm{min}}$ still reduces under fixed-rate relaying leading to weak convergence; i.e., as the number of hops grows, the $\min$ operation on the channel powers reduces both the mean and variance of the end-to-end mutual information while the loss in the mean is more than compensated by the reduction in path loss as per-hop distances become shorter. 

We note that in both fixed-rate and rate-adaptive multihop relaying, the enhancement in energy efficiency and end-to-end link reliability comes at a cost in terms of loss in spectral efficiency, as reflected through the wideband slope $S_0$, which decreases inversely
proportionally with spatial reuse separation $K$ (recall that $2 \leq K \leq N$). However, it should be emphasized that in comparison with no spatial reuse, the wideband slope improves significantly; justifying the spectral efficiency advantages of multihop routing techniques with spatial reuse in the wideband regime, especially in light of the earlier result in \cite{Oyman06b} suggesting that $S_0 = 2/N$ in quasi-static fading linear multihop networks with no spatial reuse.

\begin{figure} [t]
\centerline{\epsfxsize=3.8in \epsffile{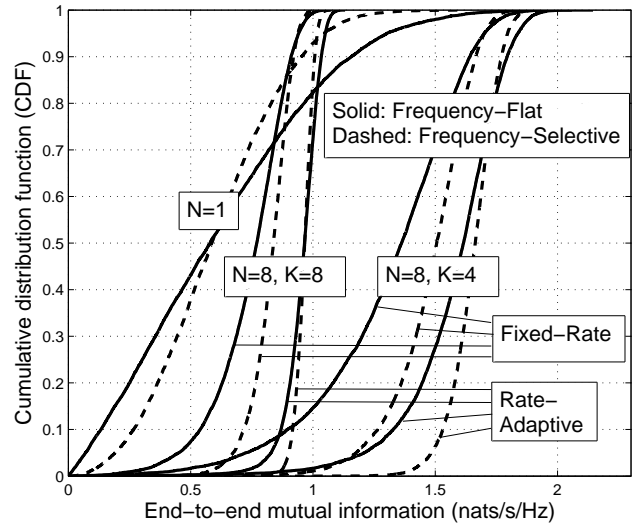}}
\caption{Cumulative distribution function (CDF) of end-to-end mutual information for fixed-rate and rate-adaptive multihop relaying schemes for various values of $N$ and $K$ in frequency-flat and frequency-selective channels.}
\label{mutual_inf}
\end{figure}

For the following numerical study, we consider multihop routing over a frequency-selective channel with $V=2$, $W=4$ as well as a frequency-flat channel with $V=W=1$. For each channel tap, the fading realization has a complex Gaussian (Ricean) distribution with mean $1/\sqrt{2}$ and variance $1/2$, under an equal-power PDP. The path-loss exponent is assumed to be $p=4$, and the average received SNR between the terminals ${\cal T}_1$ and ${\cal T}_{N+1}$ is normalized to $0\,\mathrm{dB}$. We plot in Fig.~\ref{mutual_inf} the cumulative distribution function (CDF) of the end-to-end mutual information for both fixed-rate and rate-adaptive multihop relaying schemes with varying number of hops $N = 1,8$ in cases of frequency-flat fading and frequency-selective fading; also considering spatial reuse separation values of $K=4,8$ when $N=8$. As predicted by our analysis, routing with spatial reuse combined with rate-adaptive relaying provides significant advantages in terms of spectral efficiency performance. With increasing number of hops, for both frequency-flat and frequency-selective channels, we observe that the CDF of mutual information sharpens around the mean, yielding significant enhancements particularly at low outage probabilities over single-hop communication due to multihop diversity gains. In other words; our results show that multihop diversity gains remain viable under frequency-selective fading; and may be combined with the inherent frequency diversity available in each link, to realize a higher overall diversity advantage. Finally, consistent with our analysis, the numerical results show that the rate of end-to-end link stabilization with multihopping is much faster with rate-adaptive relaying than with fixed-rate relaying. 

\section{Conclusions}

This paper presented analytical and empirical results to show the realizability of the multihop diversity advantages in the cases of fixed-rate and rate-adaptive routing with spatial reuse for wideband OFDM systems under wireless channel effects such as path-loss and quasi-static frequency-selective multipath fading. These contributions demonstrate the applicability of the multihop diversity phenomenon for general channel models and routing protocols beyond what was reported earlier in \cite{Oyman06b} and show that this phenomenon can be exploited in designing multihop routing protocols to simultaneously enhance the end-to-end link reliability, energy efficiency and spectral efficiency of OFDM-based wideband mesh networks. 
   
\begin{footnotesize}
\renewcommand{\baselinestretch}{0.2}
\bibliographystyle{IEEE}

\end{footnotesize}


\end{document}